\pdfoutput=1

\documentclass[runningheads,final]{llncs}
\usepackage[english]{babel}
\usepackage[utf8x]{inputenc}
\usepackage{mathptmx,graphicx}
\usepackage{microtype}
\usepackage{color}
\usepackage{array}
\usepackage[normalem]{ulem}

\usepackage{hyperref}
\hypersetup{%
  unicode,
  colorlinks=false,
  pdftitle={Math Indexer and Searcher Web Interface:
      Towards Fulfillment of Mathematicians' Information Needs},
  pdfauthor={Martin Líška, Petr Sojka, Michal Růžička},%
  pdfkeywords={WebMIaS, MIaS, MathML, indexing, search, canonical MathML, EuDML,
      digital libraries, information systems, information retrieval,
      mathematical content search, math indexing and retrieval,
      document ranking of mathematical papers, text mining,
      DML-CZ, Digital mathematics library projects, semantics,
      semantic annotation, similarity math search, similarity search},
  pdfsubject={Paper preprint for CICM 2014, DML track, Aveiro, Portugal, Jul 
      7-11, 2014 to be published by Springer: S.M. Watt et al. (Eds.): CICM 2014, 
      LNAI 8543, pp. 444-448, Springer International Publishing Switzerland 2014.},
}

\let\stress\emph

\begin{document}
\pagestyle{headings}  % switches on printing of running heads
\mainmatter           % start of the contributions
\title{Math Indexer and Searcher Web Interface%
       \thanks{The final publication is available at http://link.springer.com/.
               Cite as:
               S.M. Watt et al. (Eds.): CICM 2014, LNAI 8543, pp.~444--448,
               Springer International Publishing Switzerland 2014.}} 
\subtitle{Towards Fulfillment of Mathematicians' Information Needs} 
\author{Martin L\'\i\v ska \and Petr Sojka\and Michal R\r u\v zi\v cka}
\institute{Masaryk University, Faculty of Informatics, 
             Botanick\'a 68a, Brno, Czech Republic\\
           \email{martin.liski@mail.muni.cz}, \email{sojka@fi.muni.cz}, 
               \email{mruzicka@mail.muni.cz}\\ 
           WWW: \url{https://mir.fi.muni.cz/}}

\maketitle            % typeset the title of the contribution
\setcounter{page}{444}

\begin{abstract}
We are designing and developing a web user interface for
digital mathematics libraries called WebMIaS.
It allows queries to be expressed by mathematicians 
through a faceted search interface. Users 
can combine standard textual autocompleted keywords
with keywords in the form of mathematical 
formulae in \LaTeX\ or MathML formats. Formulae
are shown rendered by the web browser on-the-fly for users' feedback.
We describe WebMIaS design principles and our experiences
deploying in the European Digital Mathematics 
Library (EuDML). We further describe the issues addressed by 
formulae canonicalization and by extending 
the MIaS indexing engine with Content MathML support.

\end{abstract}

\section{The Need for a Math-Aware Search Interface}

Scalable search facilities now have the status of killer
 application on the web and are in high demand
among the users of digital mathematics libraries (DML). 
There are some papers in DMLs which contain more formulae than words.
With this in mind, we are designing and implementing 
the math-aware search engine, Math Indexer and Search 
(MIaS)~\cite{dml:doceng2011SojkaLiska} 
supporting a presentation form of mathematics, since the vast majority of
scholarly literature in math has only been available in optically recognized 
presentation formats.

MIaS has been developed primarily for use in 
EuDML~\cite{dml:borbinhaetal2011}. 
Since there is no established math-aware user interface, 
we were faced with the task of designing and implementing one. 
To gain acceptance across the wider community of potential DML users,
the main design goal was ease of use. Having the entry barrier
as low as possible is important for attracting new users.

The only available formulae search which does not have format of 
sources of documents under control was 
\href{http://www.latexsearch.com/}{\texttt{LaTeXsearch.com}} 
interface by Springer. It allows only one \LaTeX\ formula as a query. 
As the same formula can be written in many ways in \TeX, 
string hashing is used to match the query with formulae in 
documents written in \LaTeX. While most
mathematicians are used to writing a query
in \LaTeX, there are problems with this approach as formulae
similarity cannot be defined as a metric on \LaTeX\ formulae
strings. Other qualities of formulae, such as their structure should be taken
into account, as well as textual phrases denoting the content sought.
Furthermore, allowing users to type longer \LaTeX\ formulae with immediate
visual feedback simplifies the use.

For EuDML, we have added on-the-fly rendering of math, as autodetected
in \LaTeX\ and MathML formats. We have added facets for searching
in different document fields~\cite{dml:Liskaetal2011}. 
Most importantly, we have had the privilege
of mining EuDML search logs for user search scenarios 
which has shown how users have striven to find the information they require.
For example, an interesting observation was 
that Content MathML has started to appear in the math search box. 
New \href{http://dlmf.nist.gov/LaTeXML/}{\LaTeX ML} converter~\cite{LaTeXML}
allows the development of new corpora of math texts 
with math representation in both Presentation and Content 
MathML, an example of which is the database available for NTCIR-10 Math 
Task~\cite{mir:ntcir-2013} (100,000 arXiv documents).
The most challenging problems have included 
the normalization of math notations coming from different sources, 
typically a typed or copy-pasted query, and heterogeneous document formats.
Development of a robust math canonicalizer emerged as 
a~must for the success of the new math search paradigm to be
supported by the DML search user interface.

\section{User Interface for Math Information Retrieval}

\begin{figure}[tb]
\centerline{\includegraphics[width=\textwidth]{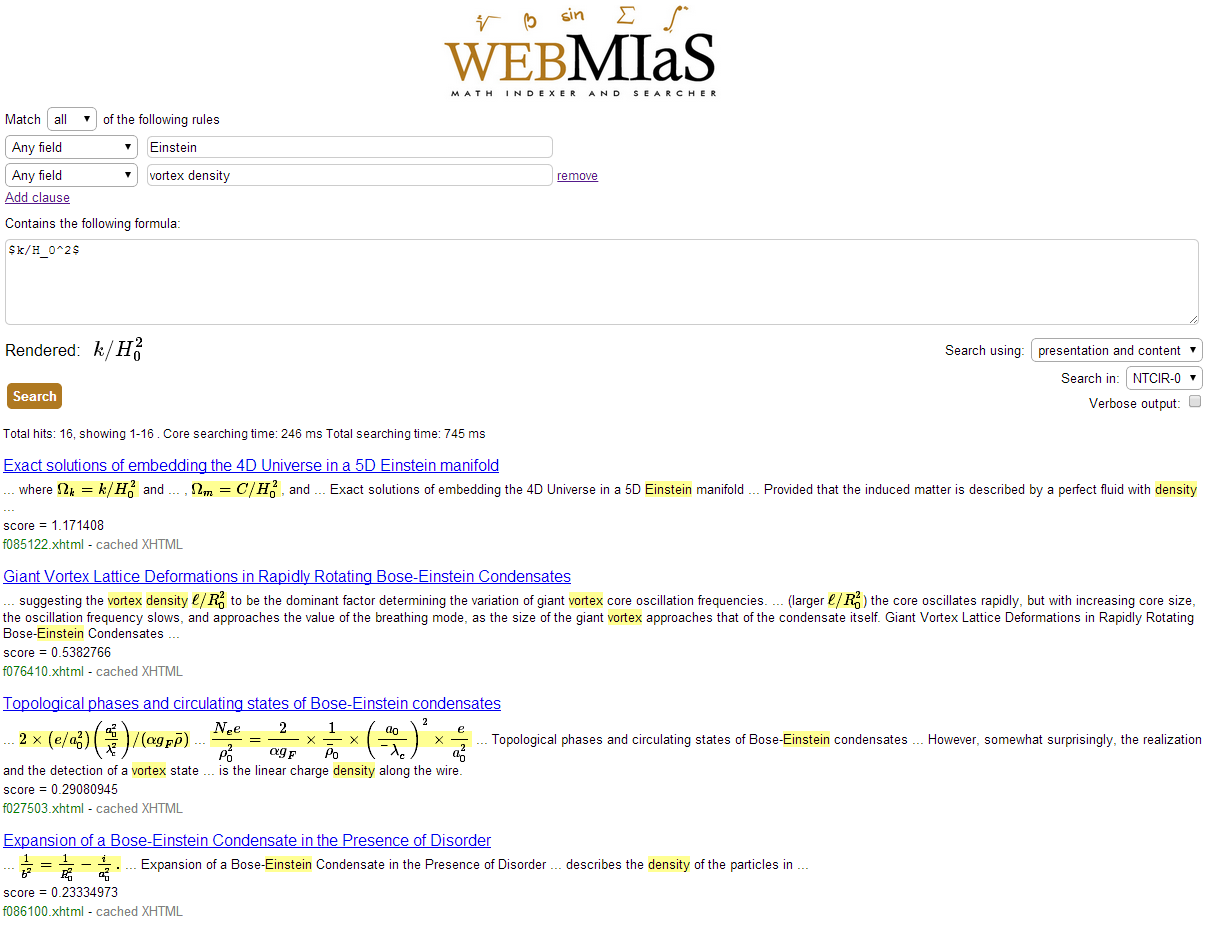}}
\caption{WebMIaS user interface}
\label{fig:webmias}
\end{figure}

Users are accustomed to forming search strategies with minimal effort
using words as queries for documents represented as bags of words.
For EuDML we have designed an advanced search 
form\hfill\\*[-.7\baselineskip]
\centerline{\url{http://eudml.org/search}} 
to allow faceted searches with one facet designed 
for inputting math formulae. On\hfill\\*[.1\baselineskip]
\centerline{\url{http://mir.fi.muni.cz/webmias}}
we maintain a link to the latest version of
the development version of WebMIaS  
to discuss possible DML users' search migration paths 
and strategies, and to get feedback from the user community. 
The WebMIaS search interface in Figure~\ref{fig:webmias} observes
several design principles and qualities:\hfill\\*[-1.2\baselineskip]
\begin{description}
\item[formulae in \TeX] Mathematicians know and use 
  compact \LaTeX\ math notation. Auto-detection of MathML 
  is also in place. To convert \LaTeX\ queries into MIaS-supported 
  MathML, we switched the converter from Tralics to \LaTeX ML, which 
  is able to convert the user input into mixed Presentation-Content MathML.
\item[on-the-fly formulae rendering] Formulae rendering allows quick feedback
  when writing the query---users know what they want when they see it. 
  Robust live rendering of copy-pasted MathML is provided means 
  of MathJax. Users are also warned when writing an invalid \TeX{} query.
\item[pop-up help] Pop-up windows inform users about the interface.
\item[domain-specific auto-completion] Frequent collocations and terms from
  the DML domain are suggested for text queries.
\item[facets] Adding facets allows natural filtering 
  (by language, author,\ldots) 
  of search results to achieve high precision.
\item[snippets with query coloring] Snippets are shown in hit lists.  
  Matched words and formulae are colored
  in the snippets for a quicker first look evaluation of the results.
\item[scoring and debugging] Scoring of computed relevance to a query is shown 
  for every hit. In the development interface, one can deduce document
  score computation.
\end{description}

Mining the EuDML and WebMIaS query search logs
reveals quite different, often contradictory
user demands. While some users prefer exact searches of visually remembered
formulae, others demand semantic specification of terms representing them. 
For example, we got a request to constrain
a search to \mbox{$E=mc^2$}, where $m$~represents mass.
The first request could be fulfilled by an exact 
\href{http://www.w3.org/TR/MathML/chapter3.html}{Presentation MathML} retrieval.
However, the latter needed semantic tagging which is usually absent in full-text XMLs
and may only be approximated by indexing disambiguated 
\href{http://www.w3.org/TR/MathML/chapter4.html}{Content MathML}.

For testing the search behaviour we indexed 
100,000 math papers from the  \mbox{NTCIR-10} Math task~\cite{NTC12}.
To formulate queries containing 
math, it emerged that new strategies will have to be developed. It can be expected that it will take some time
before math search users learn them. To find the balance between 
word and math search and between exact, proximity and subtree
search, several MIaS indexing parameters have to be set. These parameters
differ from collection to collection. Currently 
set parameters are being evaluated for the 
current collection.

As MathML created by different content generators such as 
\href{http://www.inftyproject.org/}{InftyReader}%
, \href{http://www-sop.inria.fr/marelle/tralics/}{Tralics}%
, or \href{http://latexml.mathweb.org/}{\LaTeX ML}
differ significantly. To prevent the rapid growth 
of the math index containing mathematical 
formulae~\cite{dml:doceng2011SojkaLiska}, 
their \stress{canonical representations}
with the same meaning need to be chosen and indexed. 
As the results produced by available normalization tools for MathML
were not reliable enough, we concluded that it was necessary to develop our own normalization
tool that would become part of both indexing and searching in the (Web)MIaS 
system~\cite{FormanekEtAl:OpenMathUIWiP2012}.

To help us to see the changes made by the normalization tool we are generating 
HTML reports of its inputs and results. Samples of these reports are available 
for a \href{http://dml.cz/handle/10338.dmlcz/107908}{\mbox{DML-CZ} paper}. The \LaTeX{} 
source code of the paper was transformed to XHTML+MathML by several tools. 
Examples of normalized MathML can be seen for Tralics and \LaTeX ML%
\footnote{\url{https://mir.fi.muni.cz/mathml-normalization/samples/}}%
.

When querying using math formulae~\cite{kamali2013querying}
one has to decide on the \stress{formulae similarity metrics} to allow 
not only exact formulae matches. Subformulae, formulae
expressed in different notation or even similar 
formulae with different variables 
need to be considered as hits with lower scores.
These metrics are used for weighting similar formulae
in documents in the same way that \stress{term frequency} is used for 
weighting standard word hits.

The link to the
WebMIaS interface provided above also shows our development tools that allow
us to debug MIaS indexing and querying. 
With the verbose output enabled, the computation process of 
document scores can be inspected to see how 
words and (sub)formulae affect the ranking.
The option to take only Presentation MathML,
only Content MathML or both into account is also available.
Even though Content MathML may give better, semantically related 
results most of the time, there are cases where visual 
fidelity is sought and Presentation MathML is preferred.

In addition to the web user interface, it is also possible to 
use WebMIaS remotely via web services: its general usage 
is described in the \href{http://www.opensearch.org}{OpenSearch
standard}.  Details of API and WebMIaS OpenSearch description document
can be found on the WebMIaS home page above. 
The web services are particularly useful for remote 
automated searching, as for example, for the purposes 
of system evaluation or for a programmable search.

\section{Conclusions and Future Work}

We have described the WebMIaS math-aware user interface designed
to allow searching DMLs with keyword and formulae faceted searching.
It has been applied in the EuDML project to the 
users' satisfaction. Further deployment of the WebMIaS
user interface is in preparation for further DMLs 
as DML-CZ or arXiv.

\subsubsection{Acknowledgements}
This work has been financed in part by the EU
through its Competitiveness and Innovation Programme (Information
and Communication Technologies Policy Support Programme, ``Open access
to scientific information'', Grant Agreement No. 250503).

\end{document}